# What do equitable physics lab groups look like in light of inchargeness?

Sophia Jeon, Cornell University, mj398@cornell.edu
Eleanor Sayre, Kansas State University, esayre@phys.ksu.edu
N.G. Holmes, Cornell University, ngholmes@cornell.edu

**Abstract:** In physics labs, students experience a wide range of equitable and inequitable interactions. We developed a methodology to characterize different lab groups in terms of their bid exchanges and inchargeness. An equitable group is one in which every student's bids are heard and acknowledged. Our analysis of equitable and inequitable groups raises questions about how inchargeness and gender interact to affect the functionality of a lab group.

## Introduction

Physics is the least representative of the hard sciences with measured performance gaps between male and female students. Persistence in physics, however, depends on much more than a students' test scores (Sax et al., 2001). Students' sense of belonging, for example, can be a contributing factor to their persistence (Lewis, Stout, Pollock, Finkelstein, & Ito, 2016). In college-level physics courses, students work collaboratively during labs, and the necessary coordination can support or hinder students' sense of belonging. Equity, in a general social context, implies that each student's voice is heard and each student has fair access to all parts of the learning environment (Esmonde, 2009). In this study, we present a methodology to characterize the equity of student lab groups through bids and inchargeness (Engle, Langer-Osuna, & McKinney de Royston, 2014).

As students collaborate in lab groups, they verbally or nonverbally acknowledge or follow-through with each other's bids. Inspired by Proposal Negotiation Units (Engle, et al., 2014), bids are statements in which individuals request that their partners consider an idea or complete a task (Table 1). The equity of the group can be evaluated by the relative numbers of bids made by each student and bids not acknowledged by the recipient. Inchargness relates primarily to the degree to which a person controls the conversation (Archibeque, et al., 2018), as related to their position in the group (Haaré & Van Langenhove, 1998). For example, a student may be in charge if they make many bids, many of their bids are acknowledged or followed-through by their partners, or they explain their work to the teaching assistant (TA). Students can use their inchargeness to make their groups more equitable, such as by encouraging their peers, or less equitable, such as by ignoring the bids of their peers.

Table 1: Descriptions of different kinds of bids included in the analysis.

| Bid Type | Description |
| --- | --- |
| Prompt | A question or statement that directs the flow of discourse (e.g., "What do you want to do?" or "Want to do that again?") |
| Command | A statement that directs lab partners' behavior (e.g., "You should probably write…") |
| Suggestion | An idea or plan put forward for consideration (e.g., "We can calculate the frequency…") |
| Opinion | A thought or judgment about the investigation (e.g., "I think we need more weights") |

## Methods

The participants were undergraduate students in a calculus-based introductory mechanics course at a major research university. The methodology was first developed by observing videos of two different lab groups in two-hour lab periods (an equitable pair and an inequitable trio). The two groups were deliberately selected such that one was subjectively interpreted as more equitable and the other as more inequitable. For each group, we transcribed the conversations and characterized them in terms of bids and acknowledgements of bids. Bids were identified as one of four types in Table 1. Acknowledgements of bids included verbal responses to the bids and gestures that indicated acknowledgement (such as nodding or following-through with the bid). Non-acknowledgments of bids included a lack of verbal response and gestures, and dismissals of the bids. We then tallied each students' bids to each other student, differentiating between bids directed to the group and bids directed to another individual. In this analysis, we excluded bids directed to the group. We generated representations as in Figure 1, which shows the ratio of the number of bids made by one student to the other and the number of bids not acknowledged by the recipient. We also identified and noted the moments when students explained their group's lab progress to the TA. We then analyzed the behavior of two more groups (an equitable trio and an inequitable pair). Here we summarize the analysis of an inequitable trio (Group 1) and an equitable trio (Group 2).

## Results

In Group 1, Peter, Bob, and Sandra work on their open-ended project at the end of the semester. We consider them to be an inequitable trio because they do not acknowledge many of each other's bids (Figure 1). More than half of Sandra's bids are not acknowledged by Peter, while all of Bob's bids are acknowledged by Peter. We infer that Sandra is not given the affordance to direct the group as much Peter and Bob. Peter seems to have the highest level of inchargeness because he not only made the highest number of acknowledged bids, but he also explained to the TA more than twice as many times as Bob and Sandra. This group is inequitable in light of inchargeness because Peter does not use his inchargeness to acknowledge Sandra's bids.

In Group 2, Charles, Janelle, and Veronica work on a pendulum lab at the start of the semester. We consider them to be an equitable trio because they acknowledge almost all of each other's bids. Out of these three students, Janelle has the highest level of inchargeness because she makes the highest number of bids that are acknowledged by the group. She also explains their progress to the TA 12 times, while Charles makes five explanations to the TA and Veronica makes none. This group is equitable in light of inchargeness because Janelle uses her inchargeness to acknowledge almost all of her partners' bids. Janelle also asks her partners questions, such as "What do you think?", in order to consider their opinions before her own bids are followed-through by the group.

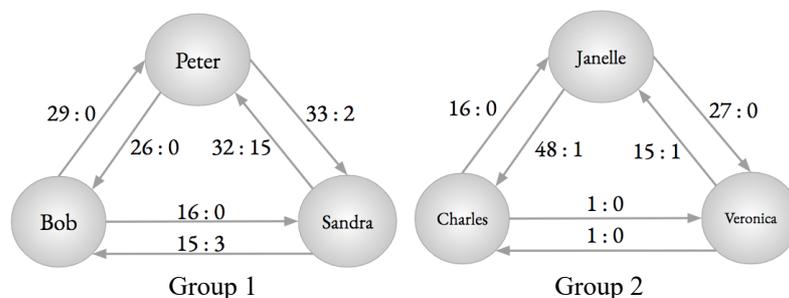

Figure 1 Bid diagram for groups showing the ratios of bids made to and not acknowledged by each student.

## Conclusions

This new methodology can be used to explore nuanced equity issues in labs. For example, the gender ratio of these groups raises questions about the impact of gender in lab group equity. Group 1 (the inequitable trio) had two men and one woman, whereas Group 2 (the equitable trio) had two women and one man. This preliminary result supports other evidence that women underperform when they are outnumbered in group work (Lewis, et al., 2016). We plan to use this methodology to investigate the equity of other mixed-gender groups and interview students about their perspectives on their lab experiences and impacts on their decisions to pursue physics.